\documentclass{PoS}

\usepackage{graphicx}
\usepackage{amsmath,amssymb}
\usepackage{capt-of}

\newcommand{\ba}{\begin{eqnarray}}
\newcommand{\ea}{\end{eqnarray}}
\newcommand{\be} {\begin{equation}}
\newcommand{\ee} {\end{equation}}

\newcommand{\order}{{\cal O}}

\title{Kaon semileptonic decay form factors with HISQ valence quarks}

\ShortTitle{$K$ semileptonic decay form factors with HISQ valence quarks}

\author{
\speaker{E.~G\'amiz}$^{a}$,
Jon~A.~Bailey$^b$,
A.~Bazavov$^c$,
C.~Bernard$^d$,
C.~Bouchard$^e$,
C.~DeTar$^f$,
D.~Du$^g$,                                                                                  
A.X.~El-Khadra$^g$,
J.~Foley$^f$,
E.D.~Freeland$^h$,
Steven~Gottlieb$^i$,
U.M.~Heller$^j$,
J.~Kim$^k$,
A.S.~Kronfeld$^l$,
J.~Laiho$^m$,
L.~Levkova$^f$,
P.B.~Mackenzie$^l$,
E.T.Neil$^l$,
M.B.~Oktay$^f$,
Si-Wei Qiu$^f$,
J.N.~Simone$^l$,
R.~Sugar$^n$,
D.~Toussaint$^k$,
R.S.~Van~de~Water$^{c,l}$,
and
Ran Zhou$^i$
 \\ \\
\llap{$^a$}
CAFPE and Departamento de F\'{\i}sica Te\'orica y del Cosmos,
Universidad de Granada, Granada, Spain\\
\llap{$^b$}Department of Physics and Astronomy, Seoul National University, Seoul, South Korea\\
\llap{$^c$}Physics Department, Brookhaven National Laboratory,\hspace*{-0.4em}
    \thanks{Operated by Brookhaven Science Associates, LLC, under
    Contract No.\ DE-AC02-98CH10886 with the United States Department
    of Energy.}~
Upton, NY, USA \\
\llap{$^d$}Department of Physics, Washington University, St.~Louis, MO, USA \\
\llap{$^e$}Department of Physics, The Ohio State University, Columbus, Ohio, USA\\
\llap{$^f$}Physics Department, University of Utah, Salt Lake City, UT, USA \\
\llap{$^g$}Physics Department, University of Illinois, Urbana, IL, USA \\
\llap{$^h$}Department of Physics, Benedictine University, Lisle, Illinois, USA\\
\llap{$^i$}Department of Physics, Indiana University, Bloomington, IN, USA \\
\llap{$^j$}American Physical Society, One Research Road, Ridge, NY, USA \\
\llap{$^k$}Department of Physics, University of Arizona, Tucson, AZ, USA \\
\llap{$^l$}Fermi National Accelerator Laboratory,\hspace*{-0.4em}
    \thanks{Operated by Fermi Research Alliance, LLC, under Contract
    No.~DE-AC02-07CH11359 with the United States Department of Energy.}~
 Batavia, IL, USA \\
\llap{$^m$}SUPA, School of Physics \& Astronomy, University of Glasgow,
Glasgow, UK\\
\llap{$^n$}Department of Physics, University of California, Santa Barbara,
CA, USA \\

E-mail: \email{megamiz@ugr.es}}
\author{Fermilab Lattice and MILC Collaborations\\
}

\abstract{
We report on the status of our kaon semileptonic form factor calculations using the highly-improved 
staggered quark (HISQ) formulation to simulate the valence fermions. We present results for the
form factor $f_+^{K \pi}(0)$ on the asqtad $N_f=2+1$ MILC configurations, discuss the
chiral-continuum extrapolation, and give a preliminary estimate of the total error. 
We also present a more preliminary set of results for the same form factor but with the 
sea quarks also simulated with the HISQ action; 
these results include data at the physical light quark masses. The improvements that we expect
to achieve with the use of the HISQ configurations and simulations at
the physical quark masses are briefly discussed.
}

\FullConference{The 30 International Symposium on Lattice Field Theory - Lattice 2012,\\
                June 24-29, 2012\\
                Cairns, Australia}

\begin{document}

\section{Introduction and methodology}

The study of exclusive semileptonic decays of $D$ and $K$ mesons provides a way of 
extracting the Cabibbo-Kobayashi-Maskawa (CKM) matrix elements $\vert V_{cd(cs)}\vert$ and 
$\vert V_{us}\vert$ with errors competitive with those obtained using other methods such 
as leptonic decays, neutrino-antineutrino interactions, or $\tau$ decays. Comparison of 
the values obtained with different methods could reveal new physics (NP) effects, and 
comparison of the shape of the form factors describing those exclusive decays with 
experiment can provide a check of the lattice methodology employed. 

Our program includes analyzing $K\to\pi l \nu$ and $D\to K(\pi)l\nu$ semileptonic decays at 
zero as well as non-zero momentum transfer. In these proceedings we focus on the 
status of the $K\to\pi l \nu$ analysis at zero momentum transfer, which one can 
combine with experimental data to extract the CKM matrix element $\vert V_{us}\vert$. 
The limiting error currently comes from the lattice 
determination of the form factors defined in (\ref{formfac})~\cite{Antonellietal2010}.  
A precise determination of $\vert V_{us}\vert$ provides stringent tests of first-row
unitarity and gives information about the scale of new physics~\cite{Cirigliano10}.

\label{sec:method}

The theory input needed to extract the CKM matrix elements from exclusive semileptonic 
widths are form factors parametrizing the corresponding hadronic matrix elements:
\ba \label{formfac}
\langle P_2\vert V^\mu\vert P_1\rangle  = f_+^{P_1P_2}(q^2)(p_{P_1}+
p_{P_2}-\Delta)^\mu +f_0^{P_1P_2}(q^2)\Delta^\mu\,,
\ea
where  $\Delta^\mu=(m_{P_1}^2-m_{P_2}^2)q^\mu/q^2$,
$q=p_{P_1}-p_{P_2}$, and $V$ is the appropriate flavor-changing vector current. 
We obtain the needed form factor $f_+^{K\pi}(0)$ using the relation 
$f_0^{K\pi}(q^2) = \frac{m_s-m_l}{m_K^2-m_\pi^2}\langle \pi\vert S
\vert K\rangle (q^2)$,
and the fact that $f_+(0)=f_0(0)$ due to the kinematic constraint. 
This method~\cite{HPQCD_Dtopi} allows us to eliminate the need for a renormalization 
factor and to extract the form factor from three-point correlation 
functions with insertion of a scalar current instead of a vector current. 

\begin{figure}[thb]
\begin{minipage}[c]{0.44\textwidth}
\includegraphics[width=1.\textwidth]{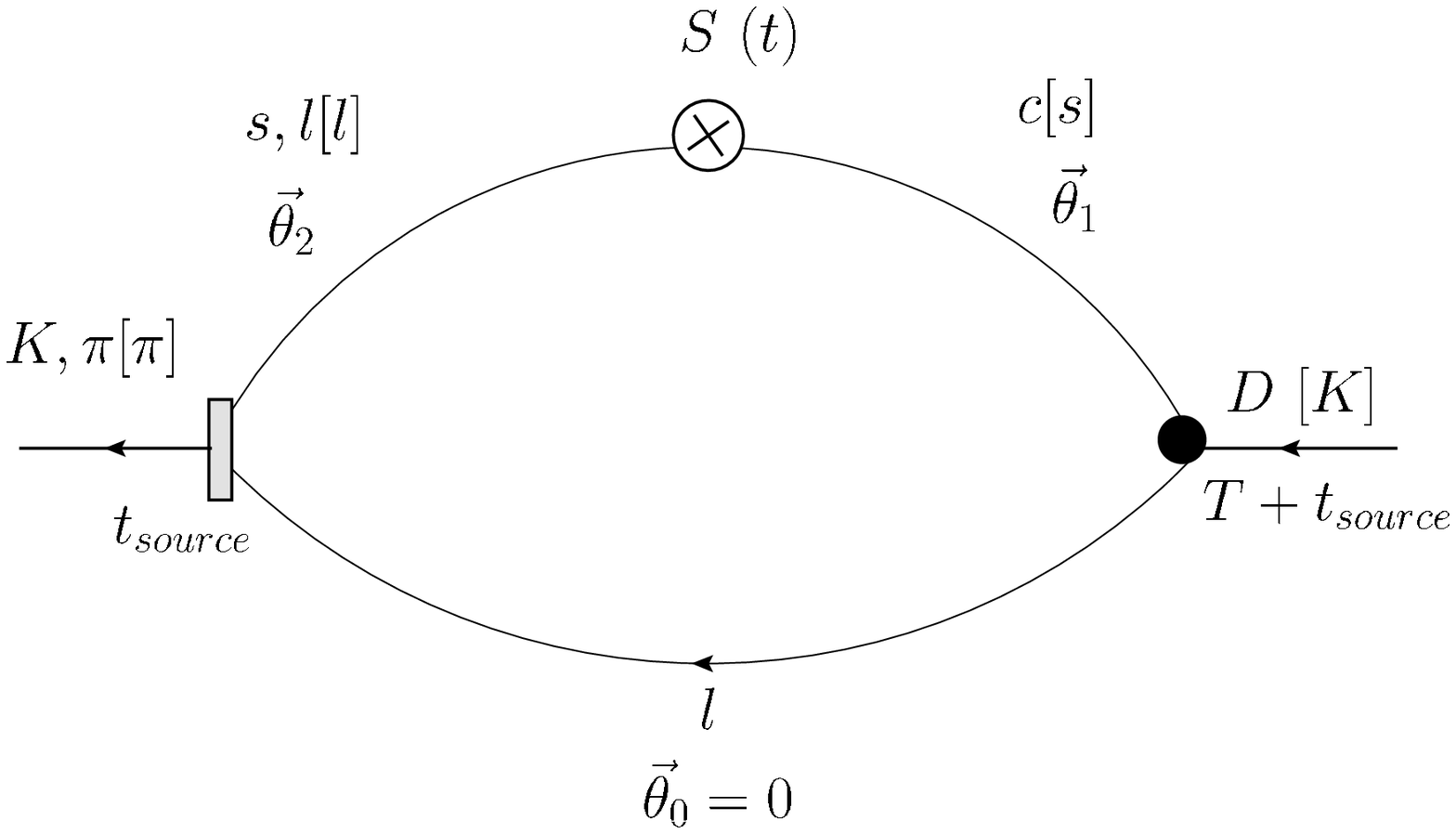}
\captionof{figure}{Structure of the 3-point functions needed to
calculate $f_{0}^{D K(\pi)}[f_0^{K\pi}]$. Light quark propagators are
generated at $t_{\rm source}$ with random wall sources. An
extended charm [strange] propagator is generated at $T+t_{{\rm source}}$. 
\label{fig:correlators}}
\vspace*{-0.5cm}
\end{minipage}
\hspace*{0.2cm}
\begin{minipage}[c]{0.54\textwidth}
\begin{tabular}{ccccccc}
\hline\hline
      $\approx a$ (fm) & $am_l/am_h$ & $N_{conf}$
& $N_{sources}$ & $N_T$ \\
\hline
$0.12$ & $0.020/0.050$ & $2052$ & $4$ & $5$ \\
       & $0.010/0.050$ & $2243$ & $4$ & $8$ \\
       & $0.07/0.050$ & $2109$ & $4$ & $5$ \\
       & $0.05/0.050$ & $2098$ & $8$ & $5$ \\
\hline
$0.09$ & $0.0124/0.031$ & $1996$ & $4$ & $5$ \\

       & $0.0062/0.031$ & $1946$ & $4$ & $5$ \\
\hline\hline
\end{tabular}
\captionof{table}{Aqstad nsembles and simulation details. $am_h$ is the nominal strange-quark mass
in the sea sector, $N_{sources}$ is the number of time sources, and $N_T$ the
number of sink-source separations. \label{tab:sim}}
\end{minipage}
\end{figure}

The momentum transfer of the three-point functions is tuned to zero or very close to 
zero using twisted boundary conditions to inject external momentum~\cite{tbc}. 
The general structure of the correlation functions  
is given in Fig.~\ref{fig:correlators}. We consider either a moving $\pi$ 
($\theta_0=\theta_1=0$ and $\theta_2\ne 0$) or a moving $K$ ($\theta_0=\theta_2=0$ 
and $\theta_1\ne 0$), for $K\to\pi l\nu$. 

\section{HISQ valence fermions and asqtad $\mathbf{N_f=2+1}$ configurations}

\label{Hisqonastad}

For the first analysis we use the asqtad $N_f=2+1$ MILC configurations and 
the HISQ action to simulate the valence quarks. The strange valence-quark mass is fixed 
to its physical value and the light valence-quark mass is fixed so 
$\frac{m_l^{{\rm val}}({\rm HISQ})}{m_s^{{\rm phys}}({\rm HISQ})}=
\frac{m_l^{{\rm sea}}({\rm asqtad})}{m_s^{{\rm phys}}({\rm asqtad})}$. 
The parameters of the ensembles included in this calculation and the details 
of the simulations we perform are collected in Table \ref{tab:sim}.

In order to extract the value of the form factor $f_+^{K\pi}(0)=f_0^{K\pi}(0)$, we 
perform a simultaneous fit of the relevant three- and two-point functions. 
The correlator fits and the multiple checks performed on their stability 
under the change of parameters, time range, number of exponentials in the fitting functions, 
and correlators included, was described in last year's conference proceedings~\cite{lat2011Ktopi}.
The conclusion is that we find it very difficult to make changes in the fitting
procedure that change the fit results outside the one sigma range.
The only change with respect to last year's fits is that in our final correlator fits we 
are including functions with the momentum injected in both the $\pi$ and in the $K$. 
The results from these combined fits for the different ensembles 
are shown in Fig.~\ref{fig:fitanalytical}. Statistical errors are very small, $\sim0.1-0.15\%$. 

\subsection{Chiral-continuum extrapolations}

\label{sec:chpt}

We need to extrapolate our results to the continuum limit and the physical light-quark 
masses, and also adjust for the mistuning of the strange sea-quark mass in the 
asqtad $N_f=2+1$ MILC configurations. Our plan is to perform these extrapolations  
using partially quenched staggered chiral perturbation theory (S$\chi$PT) at 
NLO plus regular continuum $\chi$PT at NNLO. Addressing staggered 
effects at NLO should be enough to achieve the sub-percent precision that we target.

In the continuum, the form factor is given by 
$f_+^{K\pi}(0) = 1 + f_2 + f_4 + f_6 + \dots\,$, 
where, according to the Ademollo-Gatto (AG) theorem, the chiral corrections 
$f_2$, $f_4$, $f_6$ ... go to zero in the $SU(3)$ limit as $(m_K^2-m_\pi^2)^2$.
This means that at NLO there are no free low-energy 
constants and $f_2$ is fixed in terms of experimental quantities. At finite lattice 
spacing, however, we would have violations of the AG theorem due to discretization effects 
in the continuum dispersion relation needed to derive the relation between $f_0^{K\pi}(0)$ 
and the correlation functions we are generating. 

The general structure of the fitting function we plan to use for our chiral fits is thus 

\vspace*{-0.5cm}
\ba\label{eq:ChPTtwoloop}
f_+^{K\pi}(0) = 1 + f_2^{PQ\,,stag.}(a) & + & C_4^{(a)}\,\left(\frac{a}{r_1}\right)^2 +
f_4^{cont.}({\rm logs}) + f_4^{cont.}(L_i)\nonumber\\&&
+ r_1^4\,(m_\pi^2-m_K^2)^2 \left[C_6'^{(1)}
+ C_6^{(a)}\,\left(\frac{a}{r_1}\right)^2\right]\,,
\vspace*{-0.7cm}
\ea
where the constants $C_4^{(a)}$, $C_6^{(a)}$, and $C_6'^{(1)}\propto C_{12} + C_{34}        
- L_5^2$ are free parameters to be fixed by the chiral fits. The $L_i's$
are the usual $\order(p^4)$ low-energy constants (LEC's), and the $C_{ij}$ are $\order(p^6)$ 
LEC's defined in \cite{ChPTp6}. The function $f_2^{PQ\,,stag.}(a)$ is the NLO 
partially quenched S$\chi$PT expression, which incorporates the dominant lattice artifacts 
from taste breaking. To a very good approximation, there are no free parameters in that function. 
The taste-splitting and the taste-violating hairpin parameters are already available for 
both the valence and the sea quarks from lattice calculations for asqtad or HISQ fermions,  
respectively. We are also in the process 
of calculating the taste-splittings for the mixed mesons, those made of one sea fermion 
and one valence fermion. The only parameter that we do not get from  
other calculations is an extra taste-violating hairpin parameter that appears due to 
the fact that we have a mixed action. We need to leave this quantity as a free parameter 
of the fit, although we expect its impact to be small. We take the NNLO contribution, 
$f_4$, from the calculation in~\cite{BT03}. 
Finally, we try to include terms proportional to $a^2$ with free parameters $C_4^{(a)}$ 
(to take into account the violations of the AG theorem at finite lattice 
spacing) and $C_6^{(a)}$ (to account for the residual $a^2$ dependence at higher orders). 

\begin{center}
\begin{figure}[thb]
\vspace*{-0.8cm}
\begin{minipage}[c]{0.48\textwidth}
\begin{center}
\vspace*{-0.8cm}
\hspace*{-0.2cm}\includegraphics[width=0.9\textwidth,angle=-90]{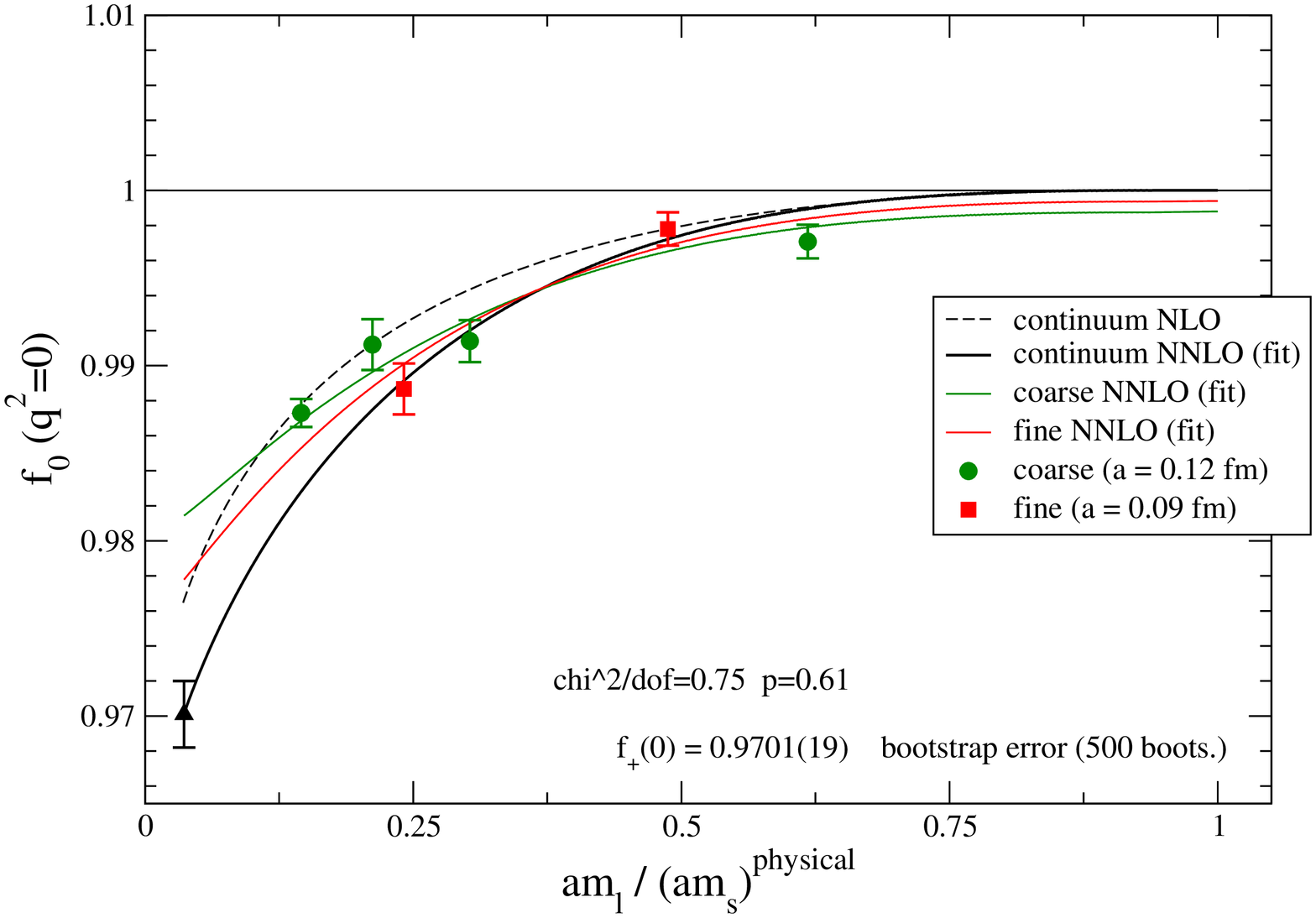}
\end{center}
\end{minipage}
\hspace*{0.3cm}
\begin{minipage}[c]{0.48\textwidth}
\begin{center}
\vspace*{0.29cm}
\includegraphics[width=1.05\textwidth]{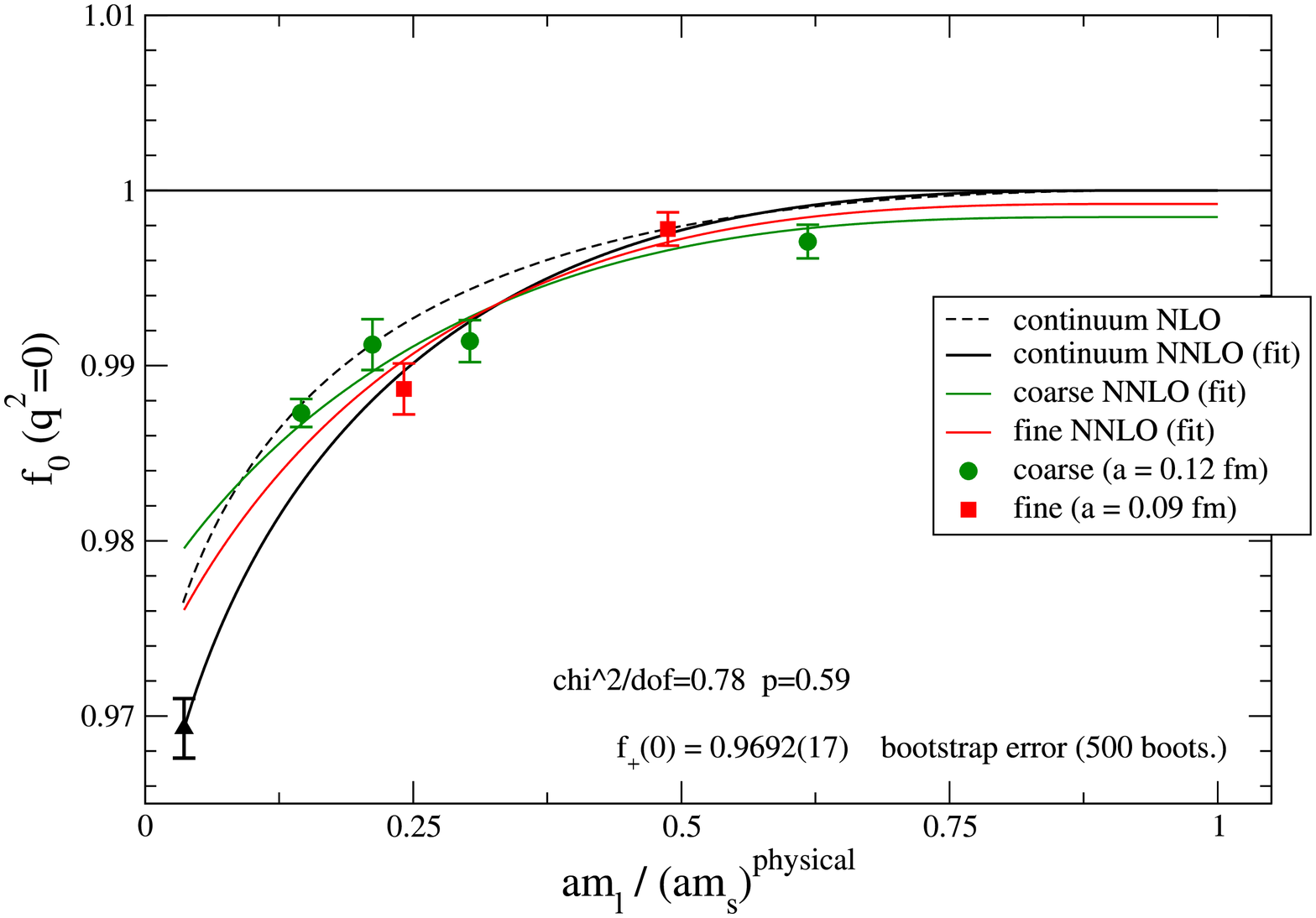}
\end{center}
\end{minipage}
\caption{Left-hand plot: example of chiral-continuum extrapolations using the fit function in
Eq.~(2.1) with $C_6^{(a)}=0$ and free $L_i$'s and priors equal to the results in 
Ref.~\cite{LisHans}. Right-hand side: example using the fit function in 
Eq.~(2.2) with $C_6^{(a)}=C_6^{(3)}=C_6^{(4)}=0$. 
Errors are statistical only, from 500 bootstrap ensembles.
\label{fig:fitanalytical}}
\end{figure}
\end{center}

Since we have not completely checked the NLO partially quenched S$\chi$PT calculation, 
here we use a simplified version without hairpin terms. We also approximate the 
taste-splittings of the mixed action mesons by the average of the sea and the valence values, 
$\Delta_{{\rm mix}}=(\Delta_{{\rm sea}}({\rm asqtad})+\Delta_{{\rm valence}}({\rm HISQ}))/2$ 
(we checked that using 
the preliminary values for the correct splitting does not change the extrapolated $f_+^{K\pi}$ 
by more than $0.1\%$). With these simplifications, 
we tried several variations of the fitting function in (\ref{eq:ChPTtwoloop}): 
fixing the LEC's $L_i$'s to their value from the global fit in~\cite{LisHans},  
fixing them to the value from the fits in~\cite{LisMILC}, or 
leaving the $L_i$'s as free parameters in the fit with prior central values equal to the 
results in \cite{LisHans} and varying the prior widths from the errors in \cite{LisHans} 
to an order of magnitude larger; including only the term proportional to $C_4^{(a)}$, the 
one proportional to $C_6^{(a)}$, or both, etc. The extrapolated value for 
$f_+^{K\pi}(0)$ has statistical errors between $0.2\%$ and $0.3\%$ in all cases and 
the different results agree with each other within one statistical $\sigma$. 
The violations of the AG theorem   
are around $0.32-0.15\%$ for $a\approx 0.12~{\rm fm}$ and $0.15-0.1\%$ for 
$a\approx 0.09~{\rm fm}$. A typical example of the fits we have performed is shown in 
the left side of Fig.~\ref{fig:fitanalytical}. 

In order to check the impact of the choice of fitting function in the extrapolation, we
have also done a number of fits replacing the NNLO continuum $\chi$PT functions by
a NNLO analytical parametrization

\vspace*{-0.6cm}
\ba\label{eq:ChPTanalytical}
f_+^{K\pi}(0) = 1 + &&f_2^{PQ\,,stag.} + C_4^{(a)} \left(\frac{a}{r_1}\right)^2 +
r_1^4(m_\pi^2-m_K^2)^2
\Big\lbrack C_6^{(1)}(r_1 m_\pi)^2 + C_6^{(2)}(r_1 m_K)^2\nonumber\\
+ && C_6^{(3)}(r_1 m_\pi)^2\ln(m_\pi^2/\mu^2) + C_6^{(4)} (r_1 m_\pi)^4 +
C_6^{(a)}\left(\frac{a}{r_1}\right)^2\Big\rbrack\,,
\ea
where $f_2^{PQ\,,stag.}$ is the same partially quenched NLO S$\chi$PT expression as in
(\ref{eq:ChPTtwoloop}) and the $C_{4}^{(a)},\, C_6^{(a)}$, and $C_6^{(i)}$ with $i=1-4$
are the free parameters in the fit. Again, we tried variations of the functional form
in (\ref{eq:ChPTanalytical}), turning on and off different NNLO and $a^2$ terms,
and parametrizing them in different ways. All the fitting functions we tried in this
category gave results within one statistical $\sigma$ of each other. 
In the right-hand side of Fig.~\ref{fig:fitanalytical} we plot an example of these fits.  
Both methods to describe NNLO contributions give results that, again,
agree within one statistical $\sigma$.

\section{Preliminary results with HISQ valence quarks and HISQ $\mathbf{N_f=2+1+1}$ configurations}

The second stage of our semileptonic decay program is the study of decays with $q^2=0$ 
for $D$ and $K$ mesons using HISQ for the valence and the sea quarks, {\it i.e.}, simulating 
on the HISQ $N_f=2+1+1$ MILC configurations.

The setup and methodology of the calculation is common to the asqtad $N_f=2+1$ calculation 
and is described in Sec.~\ref{sec:method}. Aside from the reduction in discretization 
errors provided by having HISQ sea quarks, the main improvement of this calculation 
respect to the asqtad $N_f=2+1$ one is that we will include ensembles with 
physical light-quark masses. The parameters of the ensembles we plan to include in this 
analysis, as well as the status of the runs, are shown in Table~\ref{tab:ensemblesHisq}. 
The plan is to have around 1000 configurations per ensemble and 4-5 different source-sink 
separations per ensemble at three different values of the lattice spacing. For the key 
ensembles we are generating data for 8 time sources, and, for the remaining ensembles, 
4 time sources. 
For the light and strange valence-quark masses we will use the physical values. 
For the charm-quark masses, we will simulate at a value equal to the sea charm-quark 
mass in addition to the current estimate of the physical 
one, to allow for later corrections of $am_c^{phys}$. 

\begin{table}[bht]
\begin{center}\begin{tabular}{ccccrcccc}
\hline\hline
      $\approx a$ (fm) & $am_l/am_h$ & $am_c$ & Volume & $N_{conf}$ available 
& $N_{sources}$ & $N_T$ & $\%$ run completed\\
\hline
0.15  & 0.035 & $0.831$ & $32^3\times 48$ & $1020$ & $8$ & $5$ & 100  \\
\hline
0.12 & $0.200$ & $0.635$ &  $24^3\times 64$ & $1053$ & $4$ & $4$ & 100  \\
     & $0.100$ & $0.628$ & $32^3\times 64$ & $1020$ & $4$ & $4$ & 0  \\
       & $0.035$ & $0.628$ & $48^3\times 64$ & $460$ & $8$ & $4$ & 50  \\
\hline
0.09   & $0.200$ & $0.440$ & $32^3\times 96$ & $1011$ & $4$ & $4$ & 0  \\
       & $0.100$ & $0.430$ & $48^3\times 96$ & $1000$ & $4$ & $4$ & 0  \\
       & $0.035$ & $0.432$ & $64^3\times 96$ & $497$ & $8$ & $4$ & 0  \\
\hline\hline
\end{tabular}\end{center}
\caption{HISQ ensembles and simulation details. $N_{sources}$ is the number of time sources, 
and $N_T$ the number of sink-source separations for which we have generated data. The number 
of configurations, $N_{conf}$  available and the status of the runs (last column) 
correspond to July 2012. \label{tab:ensemblesHisq}}
\end{table}

\begin{center}
\begin{figure}[thb]
\vspace*{-0.3cm}
\begin{minipage}[c]{0.48\textwidth}
\begin{center}
\vspace*{-3.5cm}                                                                              
\hspace*{-0.2cm}\includegraphics[width=0.8\textwidth,angle=-90]{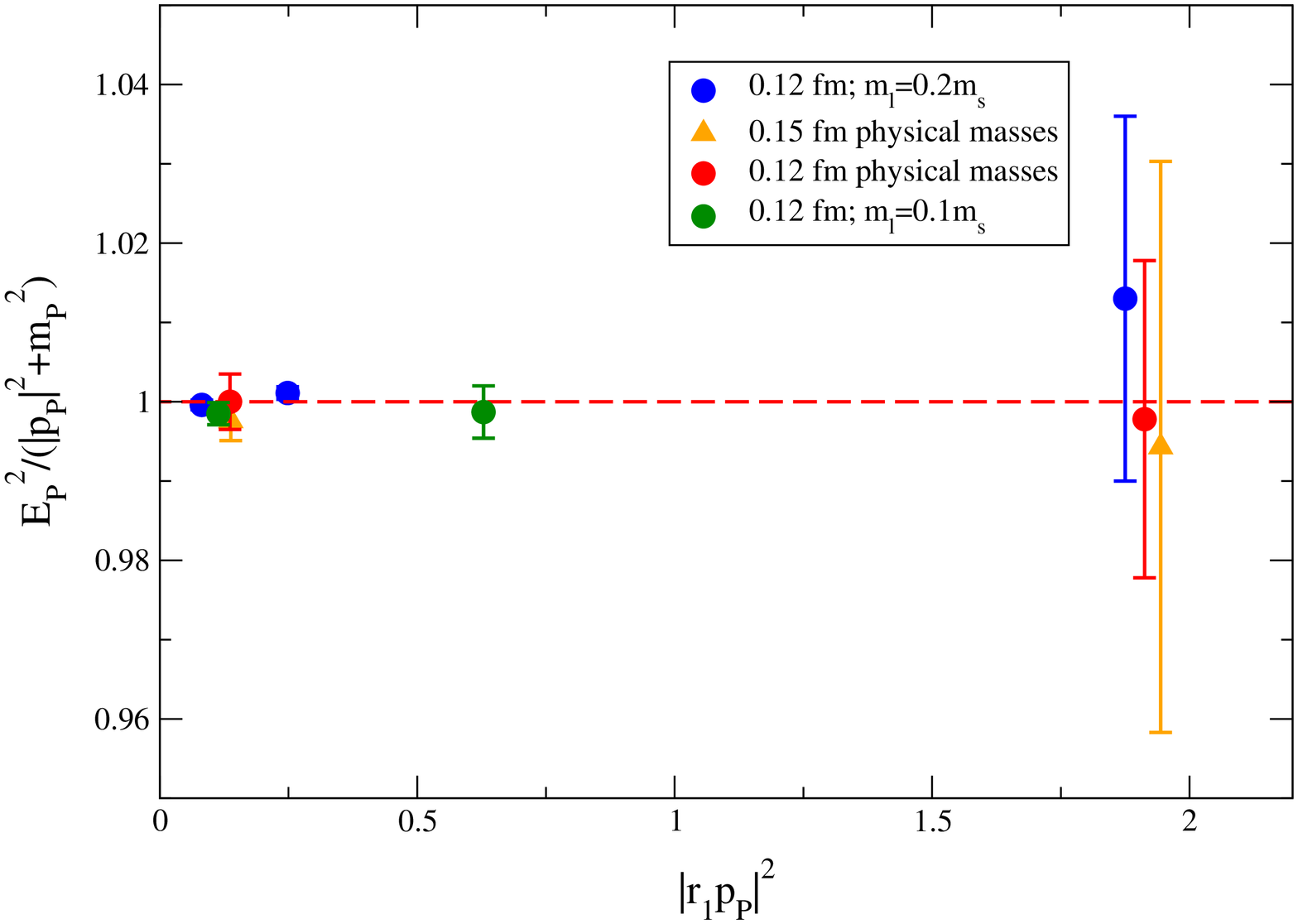}
\end{center}
\captionof{figure}{Deviation of our data from the continuum dispersion relation prediction.
\label{fig:dispersion}}
\vspace*{-0.9cm}
\end{minipage}
\hspace*{0.3cm}
\begin{minipage}[c]{0.48\textwidth}
\begin{center}

\vspace*{-1.4cm}
\hspace*{-0.3cm}\includegraphics[width=0.9\textwidth,angle=-90]{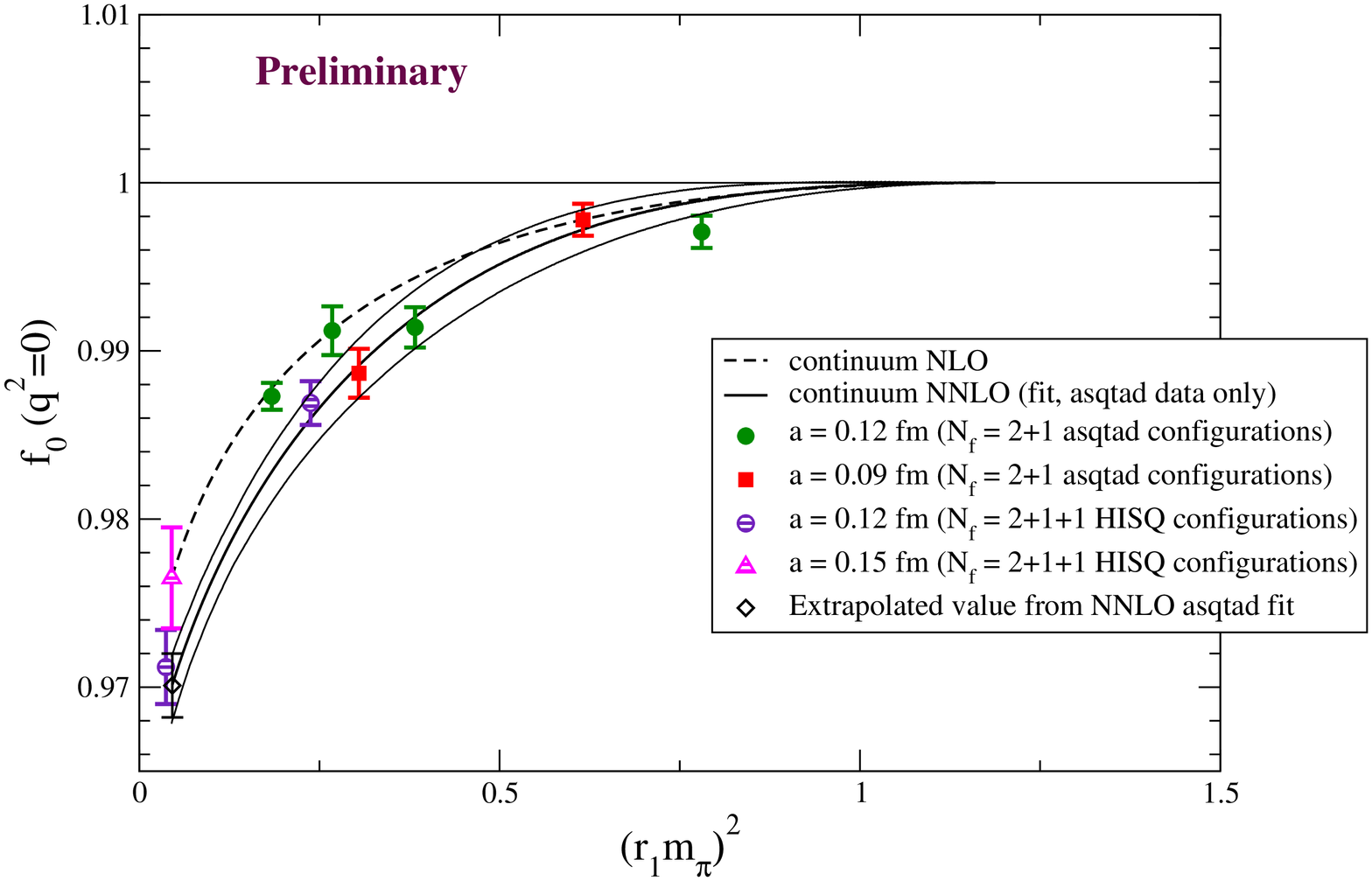}
\end{center}
\vspace*{-0.2cm}
\captionof{figure}{Form factor $f_+^{K\pi}(0)$ as a function of the $\pi$ mass from the 
asqtad $N_f=2+1$ and the HISQ $N_f=2+1+1$ calculations, together with the results from a 
fit to asqtad $N_f=2+1$ alone, also plotted on the left-hand side of 
Fig.~2. \label{fig:dataHisq}}
\end{minipage}
\vspace*{-0.5cm}
\end{figure}
\end{center}

In these runs we generate 
the correlation functions needed for the calculation of the form factor 
at zero momentum transfer for both $K\to\pi l\nu$ and $D\to K(\pi) l \nu$. 
The energies of the pions and kaons generated on those ensembles show very 
little deviation from the continuum dispersion relation, see Fig.~\ref{fig:dispersion}. 
The points with larger errors in that plot correspond to the energies needed to inject 
momentum in a $K$ to get $q^2=0$ in $K\to\pi l\nu$ when 
we have physical light-quark masses. 
Thus, for physical light-quark masses, moving pions will give us 
significantly smaller statistical errors than moving kaons in $K\to\pi l\nu$ decays. 

We fit the correlation functions generated on the HISQ configurations for the $K\to\pi l \nu$ 
decays using the same fitting functions and following the same strategy described in 
Sec.~\ref{sec:method}. The preliminary results from these fits are shown in 
Fig.~\ref{fig:dataHisq}, together with the data generated on the asqtad configurations and 
the results from the fit to the asqtad data alone (left plot in 
Fig.~\ref{fig:fitanalytical}). 
The $N_f=2+1+1$ HISQ data are very close to the continuum line obtained by 
fitting the $N_f=2+1$ asqtad data. In particular, the point corresponding to the ensemble 
with physical quark masses and $a\approx0.12~{\rm fm}$ lies right on top of the extrapolated 
value we got from the asqtad fit. This seems to indicate that the discretization effects 
in the HISQ data are going to be smaller than in the asqtad data, as expected. 
The statistical error of the physical mass point is $\sim 0.2\%$, larger than the 
$0.1-0.15\%$ error we got for larger masses, but of the same order as the extrapolated 
value. 

\section{Conclusions}

We have nearly completed the calculation of $f_+^{K\pi}(0)$ at two different values of the 
lattice spacing using the asqtad $N_f=2+1$ MILC configurations. The last step towards 
finishing the calculation is checking the NLO partially quenched S$\chi$PT expressions 
and completing the error budget. We estimate that the total error is going to be 
between $0.35-0.5\%$, dominated by the statistical and extrapolation errors ($0.2-0.3\%$) 
and the uncertainty associated with the deviation of $am_s^{sea}$ from the physical value 
($\sim 0.2\%$). We are also investigating the impact of subleading errors, such as finite 
volume effects. The total error will be competitive with current state-of-the-art 
calculations~\cite{RBC10,ETMC09}.

The dominant two errors in the calculation on the asqtad $N_f=2+1$ configurations 
will be reduced in the next step of our program namely, the calculation on the HISQ 
$N_f=2+1+1$ configurations, for which we have shown preliminary 
results here. Having data at the physical quark masses will reduce 
the statistical and extrapolation errors, as well as the one associated with the choice 
of chiral fitting function. Discretization errors are also considerably smaller for the 
HISQ action than for the asqtad action, as explicitly seen in Fig.~\ref{fig:dataHisq}. 
Finally, the strange sea-quark masses are much better tuned on the HISQ ensembles, and 
we are including the effects of the dynamical charm-quark.

\section{Acknowledgments}
We thank Johan Bijnens for making his NLO partially quenched $\chi$PT and NNLO 
full QCD $\chi$PT codes available to us.  
Computations for this work were carried out with resources provided by the
USQCD Collaboration and the Argonne Leadership Computing Facility, 
the National Energy Research Scientific Computing Center, and
the Los Alamos National Laboratory, which are funded by the Office of Science of
the U.S. Department of Energy; and with resources provided by the National
Institute for Computational Science,   
the Pittsburgh Supercomputer Center, the San Diego Supercomputer Center, 
and the Texas Advanced Computing Center, which are funded
through the National Science Foundation's Teragrid/XSEDE Program.
This work was supported in part by the MICINN (Spain) under grant FPA2010-16696
and \emph{Ram\'on y Cajal} program (E.G.), Junta de Andaluc\'{\i}a (Spain) under
grants FQM-101, FQM-330, and FQM-6552 (E.G.),
European Commission under Grant No. PCIG10-GA-2011-303781 (E.G.), 
by the U.S. Department of Energy under Grant No. DE-FG02-91ER40677 (A.X.E.) and 
DE-FG02-91ER40628 (C.W.B.), 
and by the U.S.\ National Science Foundation under grants PHY0757333 and PHY1067881 
(C.D.).

\end{document}